\documentclass[twocolumn]{aastex63} 

\newcommand{\hadot}{H$\alpha$ Dot}
\newcommand{\hadots}{H$\alpha$ Dots}

\accepted{\today}

\submitjournal{Astrophysical Journal Supplement Series}

\begin{document}

\title{\bf{The H$\alpha$ Dots Survey. IV. A Fourth List of Faint Emission--Line Objects}}


\author{Joseph D. Watkins}
\affiliation{Department of Astronomy, Indiana University, 727 East Third Street, Bloomington, IN 47405, USA}

\author[0000-0001-8483-603X]{John J. Salzer}
\affiliation{Department of Astronomy, Indiana University, 727 East Third Street, Bloomington, IN 47405, USA}

\author{Angela Van Sistine}
\affiliation{Department of Astronomy, Indiana University, 727 East Third Street, Bloomington, IN 47405, USA}
\affiliation{Center for Gravitation, Cosmology, and Astrophysics, University of Wisconsin-Milwaukee, 3135 N Maryland Ave 
Milwaukee, Wisconsin 53211, USA}

\author{Ana Hayslip}
\affiliation{Lowell Observatory, 1400 W. Mars Hill Rd. Flagstaff. Arizona. 86001. USA.}

\author[0000-0002-3744-6271]{Eric Hoar}
\affiliation{School of Materials Science \& Engineering, Georgia Institute of Technology, Atlanta, GA 30332, USA}

\author[0000-0001-7337-5936]{Rayna Rampalli}
\affiliation{Department of Physics and Astronomy, Dartmouth College, Hanover, NH 03755, USA}

\begin{abstract}

We present the fourth catalog of serendipitously discovered compact extragalactic emission-line sources  -- H$\alpha$ Dots.  A total of 454 newly discovered objects are included in the current survey list.  These objects have been detected in searches of moderately deep narrow-band images acquired for the ALFALFA H$\alpha$ project \citep{vansistine16}.  The catalog of H$\alpha$ Dots presented in the current paper was derived from searches carried out using ALFALFA H$\alpha$ images obtained with the KPNO 2.1~m telescope.  This results in a substantially deeper sample of Dots compared to our previous lists, which were all discovered in images taken with the WIYN 0.9~m telescope.  The median R-band magnitude of the current catalog is 21.59, more than 1.6 magnitudes fainter than the median for the 0.9~m sample (factor of 4.4$\times$ fainter).  Likewise, the median emission-line flux of the detected sources is a factor of 4.3$\times$ fainter.  The line-flux completeness limit of the current sample is approximately 3 $\times$ 10$^{-16}$ erg s$^{-1}$ cm$^{-2}$.  We present accurate coordinates, apparent magnitudes and narrow-band line fluxes for each object in the sample.  Unlike our previous lists of H$\alpha$ Dots, the current sample does not include follow-up spectroscopy.
\end{abstract}


\section{Introduction} \label{sec:intro}

We present the latest installment of the \hadots{} survey.   \hadots{} are compact emission-line galaxies (ELGs) discovered in narrow-band images \citep[e.g., ][]{hadots2}.  The nature of our selection method results in catalogs of strong-lined objects that identify very specific types of star-forming and active galaxies detected via a number of different emission lines.  In particular, the \hadots{} detected via the H$\alpha$ line are all dwarf star-forming galaxies (including many blue compact dwarf (BCD) galaxies) or outlying \ion{H}{2} regions in nearby spiral galaxies, while those detected via the [\ion{O}{3}]$\lambda$5007 line are typically Green Pea galaxies \citep[e.g., ][]{gp, brunker2020} or Seyfert 2 galaxies \citep{hadots2}.  The \hadots{} survey also detects high redshift QSOs via one of several UV emission lines (e.g., \ion{Mg}{2} $\lambda$2798, \ion{C}{4} $\lambda$1549 or Ly$\alpha$).

The \hadots{} survey is carried out using images acquired for the ALFALFA H$\alpha$ project \citep[][hereafter AHA]{vansistine16}.  The goal of the AHA project was to measure accurately the star-formation rate density in the local universe by obtaining H$\alpha$ images of a volume-limited sample of \ion{H}{1}-selected galaxies from the ALFALFA survey \citep{giovanelli2005, haynes2011, haynes2018}.  The \hadots{} project originated with the serendipitous discovery of point-like emission-line sources located in the narrow-band AHA images \citep[][hereafter K12]{kellar12}.  The initial discovery was made by an undergraduate student who was processing early AHA data in collaboration with the senior author.   All subsequent searches for \hadots{} in the AHA images have been carried out exclusively by undergraduate students.

Previous H$\alpha$ Dots survey lists were derived from AHA images taken with the WIYN 0.9~m telescope \citep[K12,][]{hadots2}.  A third list of \hadots{} detected using WIYN 0.9 m images is in preparation.  The current survey catalog has been created by analyzing narrow-band images obtained with the Kitt Peak National Observatory (KPNO)\footnote{The Kitt Peak National Observatory is part of the National Optical-Infrared Research Laboratory (NOIRLab) . NOIRLab is operated by the Association of Universities for Research in Astronomy (AURA) under a cooperative agreement with the National Science Foundation.} 2.1 m telescope.  As we show in the subsequent sections of this paper, the use of a larger telescope naturally results in a sample of ELGs that reaches to substantially fainter flux levels compared to the previous survey lists.

Our paper is organized as follows: Section 2 describes the observational data and the preliminary data processing steps, while Section 3 details our selection methodology.   The latter follows closely the methods adopted in the previous \hadots{} papers.  Section 4 presents our new list of \hadots{} and provides an overview of the properties of the samples using the available survey data.  Section 5 presents the limited spectroscopic information available in the literature for the newly cataloged \hadots{} and discusses potential applications of this deep sample of ELGs once follow-up spectra are obtained.  Section 6 summarizes the main results of this study.

\section{Observational Data} \label{sec:background}

The current study analyzes all ten observing runs of AHA data obtained with the KPNO 2.1~m telescope.

\begin{figure}[t]
\epsscale{1.2}
\plotone{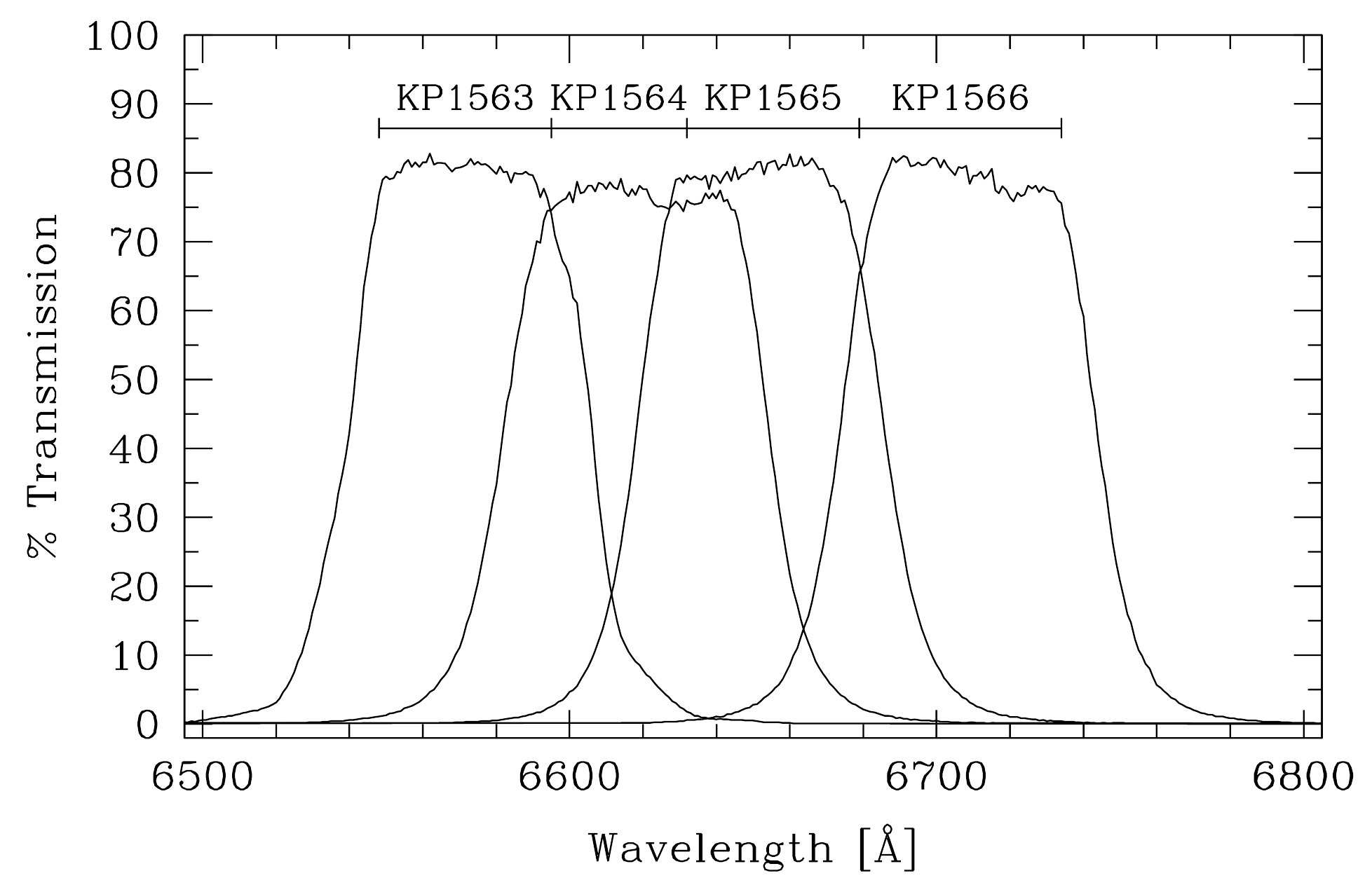}
\caption{\footnotesize The filter transmission curves for the narrow-band filter set used for the current survey. The filters are designated in the survey catalog (Table \ref{tab:catalog}) as filters 1 through 4: filter 1 = KP1563, filter 2 = KP1564, filter 3 = KP1565, filter 4 = KP1566.  The vast majority of the AHA targets observed with the 2.1~m telescope were observed in filters 3 and 4.}
\label{fig:oldhafilts}
\end{figure}

Images were taken using three different CCD detectors during the course of the AHA project: T2KB, with a FOV of 10 by 10 arcmin, and STA2 and STA3, which both had FOVs of 10 by 6.6 arcmin. All three CCDs had a pixel scale of 0.30 arcsec/pixel. Both continuum $R$-band images and narrow-band images were acquired, the latter using the KPNO narrow-band filter set (see Figure \ref{fig:oldhafilts}).  Three of the four narrow-band filters were used in the observational data set analyzed in this paper (KP1564, KP1565 and KP1566), although the majority of the AHA targets observed with the KPNO 2.1~m telescope are located in the two higher redshift filters (KP1565 and KP1566). For each target AHA galaxy, three images were obtained. First, a 15 minute narrow-band image was taken at the beginning of the observing sequence, then a 3 minute $R$-band image, followed by another 15 minute narrow-band image.

The AHA project required accurate flux calibration for the narrow-band data.  Multiple observations of spectrophotometric standard stars were acquired each night of the program.   If the photometric conditions were at all dubious on any night where data were obtained, fields taken on those nights were flagged and re-observed on a later night using short ``post-calibration" observations.   All narrow-band fluxes measured from AHA images have the zero-point of the flux scale measured to better than 2\% accuracy (and usually $\sim$1\%).  Full details are included in \citet{vansistine16}.  Because of the careful approach to flux calibration employed by the AHA program, the narrow-band fluxes measured for the \hadots{} should be quite accuarate.

Standard corrections for instrumental signatures were performed on each image. This included overscan correction, mean bias subtraction, flat field correction, and bad pixel cleaning (see K12 for details).  All preliminary image processing utilized the Image Reduction and Analysis Facility (IRAF).  
Cosmic rays were removed from the images using the L.A. Cosmic script \citep{vandokkum01}. 

A software pipeline was developed to process the AHA images; details of this pipeline are given in AHA. First, the code aligns the $R$-band and two narrow-band images to a common center and applies an astrometric solution to each image. They are then Gaussian filtered to a common stellar FWHM, after which all images have their fluxes scaled to the flux level of the first narrow-band image using bright, unsaturated stars in the field.  The scaled $R$-band image is subtracted from each narrow-band image, producing the continuum-subtracted images. Finally, the two continuum-subtracted images are added together to get the combined continuum-subtracted image.

\begin{figure*}[t]
\centering
\includegraphics[scale = 0.30]{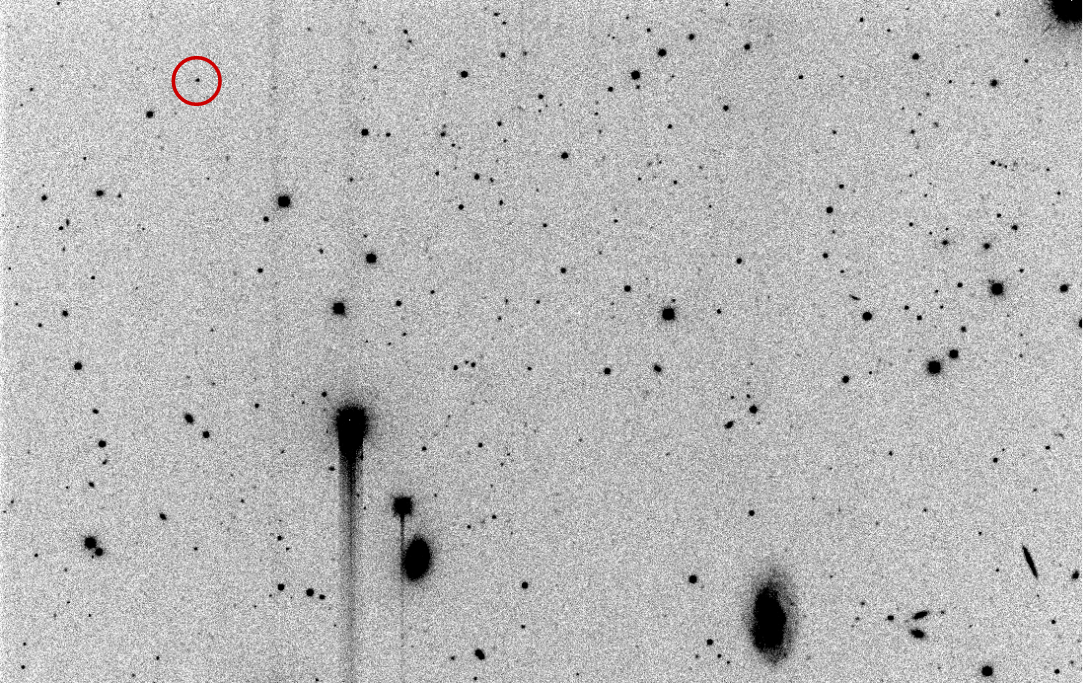} \\
\includegraphics[scale = 0.305]{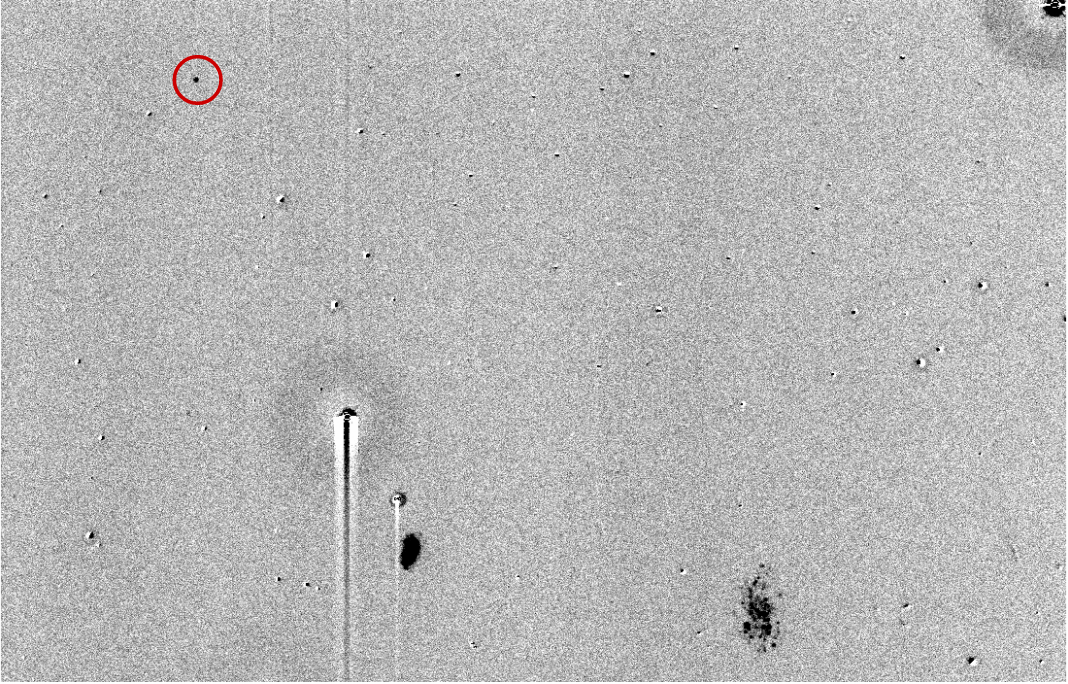}
\caption{\footnotesize  H$\alpha$ Dot 1010 is located in the upper-left portion of the field of AGC 330247. The upper image is the $R$-band continuum image, and the lower image is the continuum-subtracted narrow-band image.  The isolated point source of residual narrow-band flux is circled in red.  It has an R magnitude of 20.67.  The target AHA galaxy, AGC 330247 = CGCG 476-011, is located in the bottom-left of the field (below the two bright stars), and is seen to possess strong H$\alpha$ emission.  The spiral galaxy in the lower right is UGC 12514; it exhibits H$\alpha$ emission from a number of disk \ion{H}{2} regions.  The total field-of-view of these images is 10.0 $\times$ 6.3 arcminutes.}
\label{fig:exHAdot1}
\end{figure*}

Examples of a continuum $R$-band image and a continuum-subtracted narrow-band image are shown in Figure~\ref{fig:exHAdot1}. The images show a cut-out of the field for AHA target AGC 330247, taken with the T2KB detector.  Circled in the upper-left corner of both images is an isolated, point-like source of residual narrow-band flux.  It is unresolved in our images and is located far from either of the two larger galaxies in this field.  This object is what we call an ``H$\alpha$ Dot.'' When viewed in SDSS color images it exhibits a greenish tint, strongly suggesting that it is an [\ion{O}{3}]-detected galaxy located at z $\sim$ 0.34.  The target AHA galaxy is located in the lower-left of this image; it is a strong H$\alpha$ emitter.   The spiral galaxy located in the lower right is UGC 12514.  It is also an ALFALFA \ion{H}{1} source, and exhibits emission from many \ion{H}{2} regions in its spiral disk.

\begin{deluxetable*}{ccccc}[]
\tabletypesize{\footnotesize} 
\tablewidth{0pt} 
\tablecaption{Observing Run Summary for KPNO 2.1~m H$\alpha$ Dot Catalog \label{tab:obsruns}}
\tablehead{ \colhead{Observing Run} & \colhead{Detector} & \colhead{Number Nights} & \colhead{Number Fields} & \colhead{H$\alpha$ Dots} }
\startdata
September 2010 & T2KB & 4 & 25 & 24  \\
November 2010 & T2KB & 5 & 45 & 26  \\
March 2011 & T2KB & 7 & 72 & 78  \\
May 2011 & T2KB & 7 & 48 & 53  \\
October 2011 & T2KB & 8 & 88 & 54  \\
March 2012 & T2KB & 9 & 87 & 89 \\
September 2012 & STA2 & 8 & 80 & 45  \\
March 2013 & T2KB & 5 & 45 & 28  \\
April 2013 & T2KB & 8 & 68 & 34  \\
March 2014 & STA3 & 5 & 53 & 23  \\
Totals &  & 66 & 611 & 454 \\
\enddata
\end{deluxetable*}

A total of 611 ALFALFA H$\alpha$ fields were observed over the course of 10 observing runs (Table \ref{tab:obsruns}). These 10 runs can be broken up into Fall sample and Spring samples. The Fall sample is approximately contained within a region between R.A. of $ 22^{h}$ to $3^{h}$ and Dec. of $+24^{\circ}$ to $+29^{\circ}$, and the Spring sample is approximately contained between R.A. of $ 7^{h} 30^{m}$ to $ 16^{h} 40^{m}$ and Dec. of $+3^{\circ}$ to $+17^{\circ}$. The images collected during the course of these ten observing runs were searched for H$\alpha$ Dots. Our methodology for detecting \hadots{} in the processed AHA narrow-band images is described in the next section.

\section{2.1~m H$\alpha$ Dot Survey: Identifying H$\alpha$ Dots} \label{sec:methodology}

For an object in the field to be considered to be an \hadot, it must satisfy two primary criteria.  First, it must have a statistically significant excess of flux in the narrow-band filter relative to the R-band flux.   Second, it must be morphologically compact.  This usually means the object is either unresolved or barely resolved in our CCD images.   The first criterion is readily quantified (see below), while the second is admittedly somewhat subjective.  In particular, it will vary from field-to-field depending on the size of the point-spread function (a.k.a. ``seeing") associated with each image.  All \hadot\ candidates are evaluated by at least two members of the project team in order to try to invoke a uniform assessment of compactness.

A software package was developed in order to automatically and systematically search for \hadots{} in the AHA images.  The software employs routines designed to identify every object in an image, compare their fluxes in the continuum and narrow-band filters, and then calculate a magnitude difference and its uncertainty for each source.  Potential candidates are reviewed by members of the \hadot\ team before the list of ELG candidates is finalized.  For more details about the software package, see K12.

The software takes as input the $R$-band and composite narrow-band images and identifies every object present in the field using a modified version of DAOFIND \citep{stetson87}.  Next the software performs photometry with a constant aperture size on each object in the scaled $R$-band and the {\it unsubtracted} narrow-band images to construct a magnitude difference, calculated as 

\begin{equation}
    \Delta m = m_{NB} - m_{R}.
\end{equation}{}

\noindent Here the magnitudes used are simple instrumental magnitudes.  Because the images being used have all been scaled to a common flux level, objects with no emission lines (e.g., most stars) will have $\Delta$m = 0.0.  Large negative values of $\Delta$m indicate an object with a significant excess of flux in  the narrow-band image.  The software also computes the ratio of the absolute value of the magnitude difference to the error in the magnitude difference, as

 \begin{equation}
    ratio = \frac{| \Delta m |}{\sigma_{\Delta m}},
\end{equation}

\noindent where $\sigma_{\Delta m}$ is generated by taking the errors associated with the $R$-band and narrow-band magnitudes and summing them in quadrature:

\begin{equation}
    \sigma_{\Delta m} = {\sqrt{\sigma_{NB}^2 + \sigma_{R}^2 }}.
\end{equation}

\noindent The $ratio$ parameter serves as a pseudo signal-to-noise (SNR) ratio.  Small values of $ratio$ represent either objects with little or no emission-line flux (small $\Delta$m) or noisy sources (large $\sigma_{\Delta m}$).

\begin{figure}[]
\begin{center}
\epsscale{1.1}
\plotone{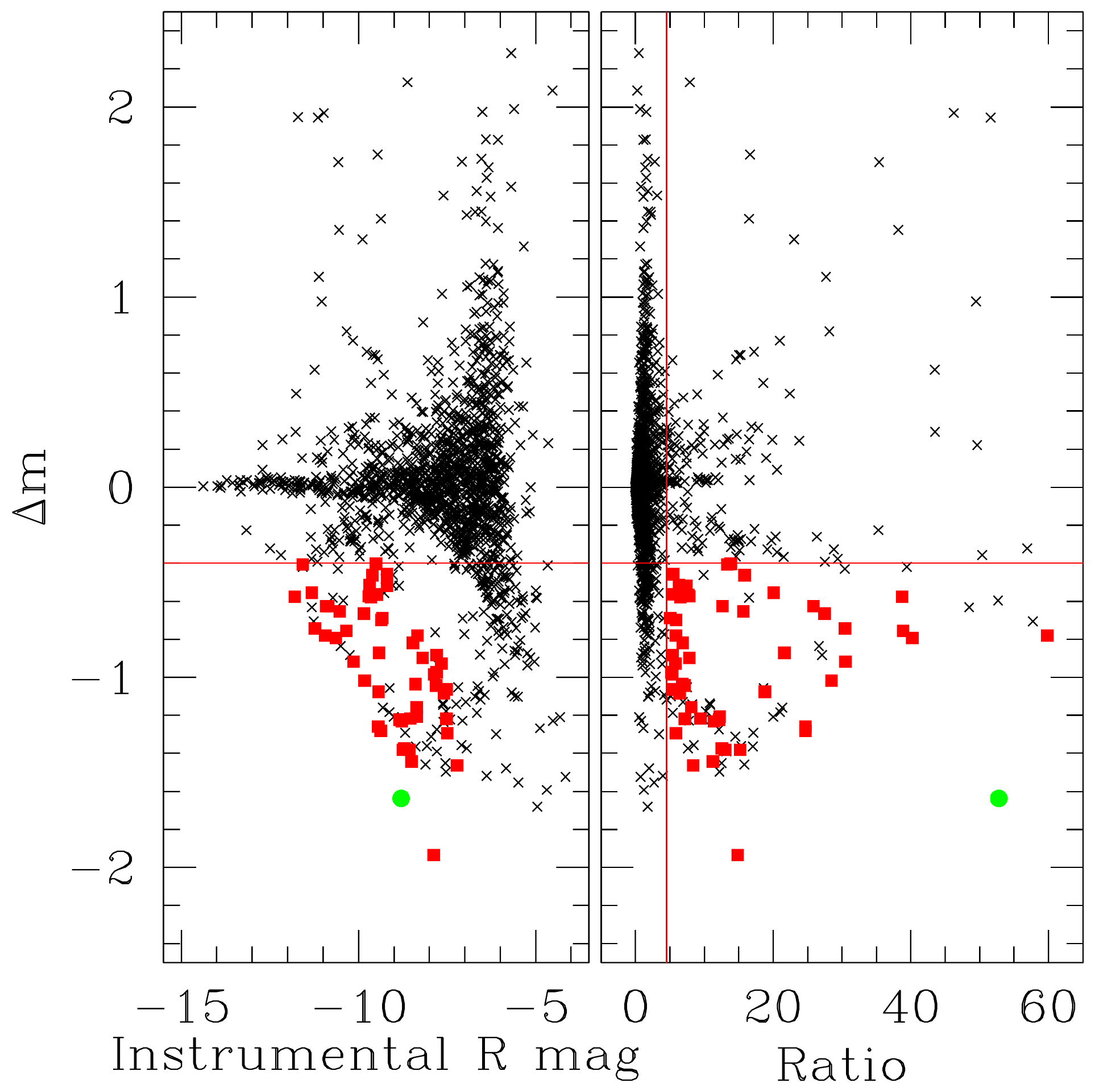}
\caption{\footnotesize  The diagnostic plot produced after using the dot-finding software on the field of AGC 330247 (see Figure~\ref{fig:exHAdot1}).  The left panel plots $\Delta m$ (eq. 1) vs. instrumental R-band magnitude.  Brighter stars are located to the left and lie along a line around $\Delta m = 0$ since the narrow-band and broad-band flux levels are normalized to a common value by our software.  Objects with a negative $\Delta m$ indicate residual narrow-band flux.  The right panel plots $\Delta m$ vs. $ratio$ (eq. 2).  Objects of interest have a large negative $\Delta m$ and large $ratio$ values, corresponding to the bottom-right quadrant of the plot.  The solid red lines indicate the limiting values for $\Delta m$ and $ratio$ for inclusion in the sample (see text).  The large number of putative detections in this field is caused by the many \ion{H}{2} regions located in AGC 330247 and UGC 12514 (Figures~\ref{fig:exHAdot1} and~\ref{fig:HIIregion}).   The \ion{H}{2} regions (n = 52) are marked in both panels with red squares, while the single \hadot{} candidate (\hadot{} 1010) is shown as a green circle.   The remaining objects in the lower right quadrant, indicated by the ``$\mathsf{X}$'' symbol, are all image artifacts that have been rejected.   We also point out the many bright (R$_{Inst}$ $<$ $-$8) sources with large {\it postive} $\Delta m$ in the upper left portion of the left panel.  These are all image artifacts caused by the long saturation ``bleed" trail from the bright star visible in Figure~\ref{fig:exHAdot1}. }
\label{fig:diagplot}
\end{center}{}
\end{figure}

For each field analyzed, the software generates a diagnostic plot (see Figure \ref{fig:diagplot} for an example).  Each ``$\mathsf{X}$'' in the plot indicates a single object in the images. The left-hand graph plots $\Delta m$ against the instrumental R-band magnitude. The bright stars in the field are clumped at $\Delta m$ = 0 on the left side of the plot; the locus of stars remains centered on zero but spreads out to larger values of $\Delta m$ for the fainter stars with large photometric errors.  Objects with a negative magnitude difference indicate more flux in the narrow-band image than in the $R$-band image. The right-hand graph plots $\Delta m$ against the $ratio$ parameter.  The vertical and horizontal lines drawn on the diagnostic plot represent the threshold values for $\Delta m$ and $ratio$ that are used to select emission-line candidates.  We set the values for inclusion in the \hadot{} sample at $\Delta m$ $<$ $-$0.4 and $ratio$ $>$ 4.5. These values were found by K12 to optimize the detection of faint objects and minimize the number of false detections.  Objects located in the lower-right quadrant of the right-hand plot of Figure~\ref{fig:diagplot} represent candidate \hadots{}. 

{}

Once the software has selected all possible candidates, these objects must be visually examined to ascertain their nature.  This verification step is essential, since the automated software and our high-quality data combine to yield numerous sources that are not true emission-line objects.    Our survey images typically include numerous sources that can lead to false detections, including uncleaned cosmic rays, saturated stars, satellite or meteor trails, and noise spikes. It is also common that star--forming regions in the AHA target are also selected by passing the $\Delta m$ and $ratio$ criteria listed above.  For example, many of the emission-line candidates in the diagnostic plot shown in Figure~\ref{fig:diagplot} are \ion{H}{2} regions in UGC 12514 (see Figures~\ref{fig:exHAdot1} and~\ref{fig:HIIregion}).  The object review process is necessary to separate these three types of detections: real \hadots{}, \ion{H}{2} regions, or image artifacts. 

\begin{figure*}[t]
\begin{center}
\includegraphics[scale = 0.42]{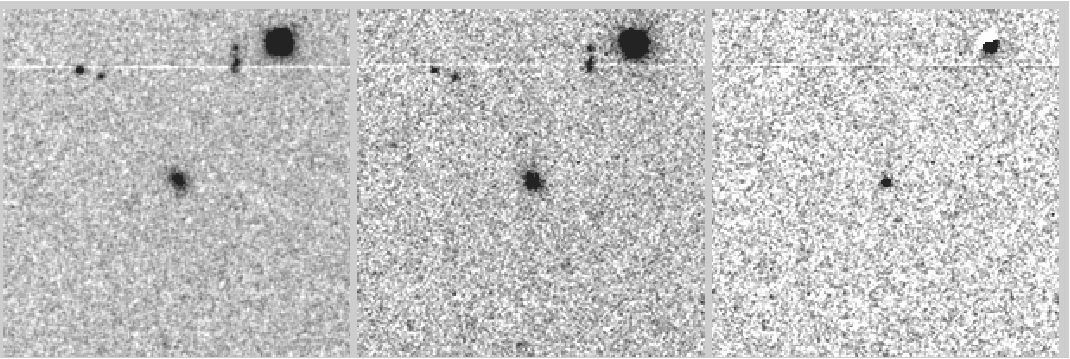}
\caption{\footnotesize Example of image cutouts used to evaluate each \hadot{} candidate.  The field is centered on \hadot{} 1447.  From left to right, these images are the $R$-band continuum image, the combined narrow-band image, and the continuum-subtracted narrow-band image. The compact appearance and significant residual flux present in the continuum-subtracted image is characteristic of an \hadot{}. Each sub-image is 200 $\times$ 200 pixels, or 60 arcseconds on a side.}
\label{fig:hadotinfield}
\end{center}{}
\end{figure*}{}

\begin{figure*}[]
\begin{center}{}
\includegraphics[scale = 1.0]{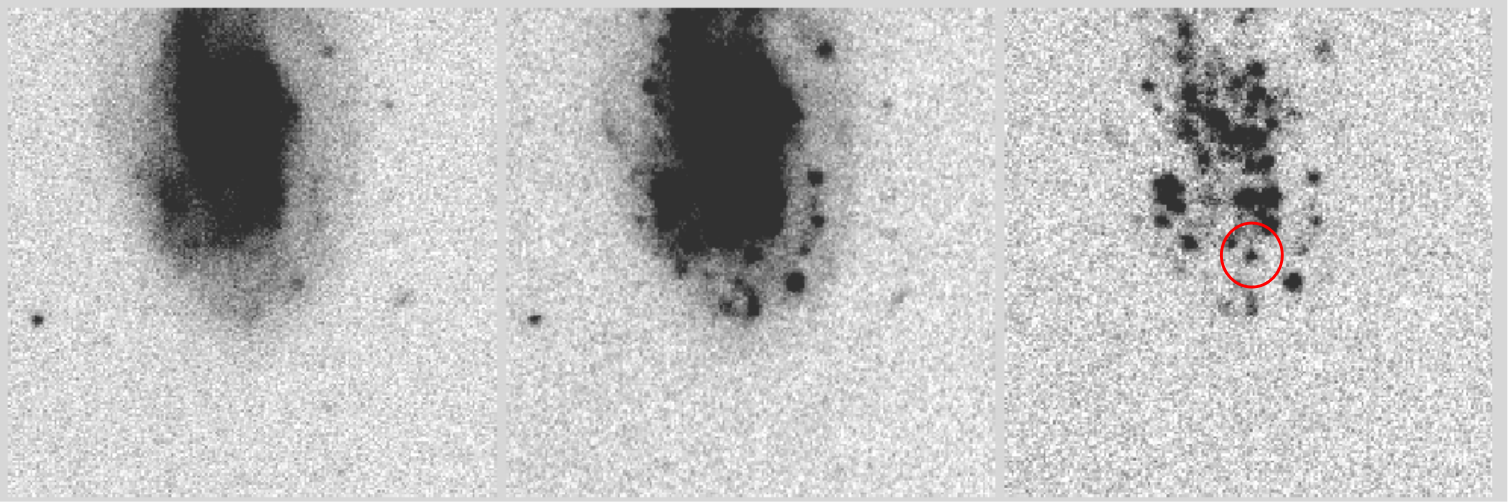}
\caption{\footnotesize This set of image cutouts shows an example where an \ion{H}{2} region in a large spiral galaxy has been detected by the dot-finding software (indicated by the red circle). The software often detects dozens of \ion{H}{2} regions in the central AHA galaxy because they surpass the $\Delta m$ and $ratio$ thresholds.   In the case of this particular galaxy (UGC 12514) a total of 37 \ion{H}{2} regions were selected by our software.  The review of all objects located in the lower right quadrant of the diagnostic diagram (e.g., Figure~\ref{fig:diagplot}) allows the user to categorize this detection properly as an \ion{H}{2} region and not an \hadot{}.}
\label{fig:HIIregion}
\end{center}{}
\end{figure*}{}

Our software produces image cut-outs for each object found in the bottom-right quadrant of the diagnostic plot.   These are three sub-images (200 $\times$ 200 pixels in size) centered on the object in question and displayed horizontally.  The cut-outs contain the object as seen in the $R$-band image, the combined narrow-band image, and the continuum-subtracted narrow-band image (see Figures \ref{fig:hadotinfield} and \ref{fig:HIIregion}).  Using the image cut-outs, the user visually examines each object located in the bottom--right quadrant of the right plot in Figure \ref{fig:diagplot} and categorizes them into one of the three object types specified above.  Objects that are classified as image artifacts are discarded.   Objects flagged as \hadots{} or HII regions are cataloged into separate lists for subsequent analysis.

An object must be compact in appearance, spatially separate from the central AHA galaxy, and contain significant emission in the narrow-band image in order to be selected as an \hadot{} candidate.   These criteria are discussed in detail in the previous \hadots{} survey papers \citep{kellar12, hadots2}.   The compactness criteria was instituted to avoid cataloging large, extended galaxies that were already known.  In particular, we did not wish to include the target AHA galaxy in our survey.   The compactness and separation requirements are somewhat subjective, although the visual checking by at least two independent team members helps to ensure a fairly uniform approach.   We note that the separation criterion does not prevent the survey from detecting isolated / outlying \ion{H}{2} regions that are associated with the AHA target galaxy.  Several examples are included in \citet{hadots2}, who emphasize that the identification of such objects requires follow-up spectroscopy. 

Examples of objects detected in our emission-line searches are shown in Figures \ref{fig:hadotinfield} and \ref{fig:HIIregion}.   The object in Figure~\ref{fig:hadotinfield} is \hadot{} 1447, which has a measured R-band magnitude of 19.87 $\pm$ 0.03.  While very compact in nature, is is seen to be resolved in the broad-band image.  This implies that the emission line detected in our narrow-band filter is most likely H$\alpha$.  Since it was observed in filter KP1566 (i.e., filter 4), its redshift is most likely to fall between 5300 and 7800 km/s (\citep{vansistine16}. This would make \hadot{} 1447 a dwarf star-forming galaxy, which is consistent with its blue appearance in the Sloan Digital Sky Survey (SDSS) color images \citep{sdss, sdss14}.  Figure~\ref{fig:HIIregion} shows a number of \ion{H}{2} regions located in the nearly face-on spiral UGC 12514 (see Figure~\ref{fig:exHAdot1}).  Large spiral galaxies like this are commonly detected multiple times by our software.  In the example shown the detected \ion{H}{2} region is the faint object located at the center of the cut-outs.  As mentioned above, these \ion{H}{2} regions are cataloged separately from the \hadots{}.  They are not discussed further as part of the current study.

After the review process is completed, objects flagged as \hadots{} are remeasured more carefully in order to obtain accurate $R$-band magnitudes and narrow-band line fluxes for each source.  The photometric calibrations derived for the AHA project are utilized to place the measurements on an accurate flux scale. 

Once the list of \hadots{} discovered in the AHA 2.1~m images was finalized, the entire catalog was crossed-matched with objects in the SDSS Data Release 14 \citep{sdss14}.  Using the coordinates for the \hadots{} obtained from our astrometric solution (see Section \ref{sec:background}), SDSS positions were retrieved for all \hadots{} that were within the SDSS footprint. We then visually compared the location given by the coordinates for each \hadot{} to verify that the query had returned the correct object.  If it returned the wrong object, the correct object was located using the SDSS navigate window, and new SDSS coordinates were obtained by centering the cursor on the object in the navigate window and reading off the corresponding Right Ascension and Declination.  Fully 83.7\% of the 2.1~m \hadots{} (380 of 454) were included in the SDSS photometric catalog.  The \hadots{} not in SDSS were either too faint (65 of 454, or 14.3\%) or are located outside of the SDSS footprint (9 or 454, or 2.0\%).  For the objects in common between the two surveys we queried SDSS DR14 again to obtain the full set of SDSS {\it ugriz} magnitudes and errors.  This information was then merged into the \hadots{} database.

\section{2.1~m H$\alpha$ Dot Survey: Results} \label{sec:discussion}

The final catalog of \hadots{} detected using the KPNO 2.1~m AHA images contains 454 newly discovered ELGs. This list of objects was arrived at after searching 611 AHA fields.   The total sky coverage of these images represents 15.494 square degrees, resulting in a surface density of new \hadots{} of 29.30 deg$^{-2}$.   This number, which is a key figure of merit for such surveys, is substantially higher than the surface densities of objects detected in the previous \hadots{} survey lists \citep[K12,][]{hadots2}.   The latter had surface densities of 5.22 and 5.24 ELGs/deg$^2$, respectively, more than a factor of 5.5$\times$ lower.

\begin{deluxetable*}{crrcccrcccc}[t]
\tabletypesize{\footnotesize}
\tablewidth{0pt}
\tablecaption{Fourth List of H$\alpha$ Dots \label{tab:catalog}}

\tablehead{
  \colhead{H$\alpha$ Dot \#} &   \colhead{RA(2000)} &   \colhead{DEC(2000)} &    \colhead{Obs. Run} &   \colhead{Filter} &  \colhead{$\Delta$m} &   \colhead{Ratio} &  \colhead{m$_R$} &  \colhead{NB Line Flux} & \colhead{SDSS r}    \\
  &  \colhead{degrees} &  \colhead{degrees} &  &  &  &  &  \colhead{mag}  & x10$^{-14}$ erg/s/cm$^2$ &    \colhead{mag}  \\
 (1)  & \colhead{(2)}  & \colhead{(3)}  & (4)  & (5)  & (6)  & \colhead{(7)}  & (8)  & (9)  & (10)
}
 \startdata
 1001 & 330.44398 &  24.18998 & Sep2010 & 4 &  -1.21 &  23.91 &  21.84 $\pm$   0.06 &    0.062 $\pm$    0.003 & 22.05 $\pm$ 0.10 \\
 1002 & 330.44587 &  24.20362 & Sep2010 & 4 &  -0.92 &  72.05 &  19.27 $\pm$   0.02 &    0.329 $\pm$    0.004 & 19.45 $\pm$ 0.02 \\
 1003 & 330.45553 &  24.21585 & Sep2010 & 4 &  -0.60 &  34.08 &  19.69 $\pm$   0.02 &    0.179 $\pm$    0.004 &   \\
 1004 & 334.14431 &  27.94493 & Sep2010 & 4 &  -1.20 &  17.60 &  21.39 $\pm$   0.04 &    0.058 $\pm$    0.004 &   \\
 1005 & 340.46256 &  27.76333 & Sep2010 & 4 &  -0.49 &  10.80 &  21.13 $\pm$   0.05 &    0.043 $\pm$    0.004 &   \\
 1006 & 340.55283 &  27.69124 & Sep2010 & 4 &  -1.25 &  18.51 &  22.13 $\pm$   0.10 &    0.059 $\pm$    0.004 & 21.87 $\pm$ 0.11 \\
 1007 & 340.59996 &  27.81721 & Sep2010 & 4 &  -1.89 &  12.94 &  22.94 $\pm$   0.18 &    0.057 $\pm$    0.004 &   \\
 1008 & 344.40155 &  26.40989 & Sep2010 & 4 &  -0.72 &  30.60 &  19.98 $\pm$   0.03 &    0.135 $\pm$    0.005 & 20.19 $\pm$ 0.03 \\
 1009 & 349.23530 &  27.93332 & Sep2010 & 4 &  -1.10 &   7.39 &  22.97 $\pm$   0.28 &    0.033 $\pm$    0.006 & 22.87 $\pm$ 0.19 \\
 1010 & 350.08029 &  26.09913 & Sep2010 & 4 &  -1.64 &  52.78 &  20.67 $\pm$   0.05 &    0.348 $\pm$    0.007 & 20.66 $\pm$ 0.04 \\
\\
 1011 & 352.82919 &  25.12090 & Sep2010 & 4 &  -0.54 &  17.84 &  20.48 $\pm$   0.03 &    0.092 $\pm$    0.006 & 20.82 $\pm$ 0.05 \\
 1012 & 355.59838 &  27.96453 & Sep2010 & 4 &  -0.42 &   7.10 &  21.37 $\pm$   0.05 &    0.026 $\pm$    0.004 & 21.81 $\pm$ 0.16 \\
 1013 & 355.70529 &  27.97838 & Sep2010 & 4 &  -1.63 & 106.59 &  19.44 $\pm$   0.02 &    0.762 $\pm$    0.008 & 19.78 $\pm$ 0.04 \\
 1014 & 355.70809 &  28.07243 & Sep2010 & 4 &  -1.87 &   7.21 &  23.75 $\pm$   0.58 &    0.017 $\pm$    0.004 &   \\
 1015 & 355.99726 &  27.15461 & Sep2010 & 4 &  -2.17 &   8.94 &  22.51 $\pm$   0.12 &    0.053 $\pm$    0.006 &   \\
 1016 & 356.00110 &  27.15570 & Sep2010 & 4 &  -1.63 &   7.17 &  22.02 $\pm$   0.11 &    0.057 $\pm$    0.006 &   \\
 1017 & 356.00053 &  27.12553 & Sep2010 & 4 &  -0.50 &  11.08 &  21.16 $\pm$   0.04 &    0.054 $\pm$    0.006 & 21.07 $\pm$ 0.09 \\
 1018 & 356.05400 &  27.19430 & Sep2010 & 4 &  -0.72 &   7.08 &  22.38 $\pm$   0.10 &    0.033 $\pm$    0.004 & 22.00 $\pm$ 0.15 \\
 1019 & 357.03115 &  24.32929 & Sep2010 & 4 &  -1.48 &   6.31 &  24.87 $\pm$   2.26 &    0.041 $\pm$    0.006 &   \\
 1020 & 357.18135 &  27.98432 & Sep2010 & 4 &  -0.41 &   4.92 &  21.38 $\pm$   0.09 &    0.017 $\pm$    0.003 & 21.43 $\pm$ 0.09 \\
\\
 1021 & 357.18551 &  27.94486 & Sep2010 & 4 &  -1.28 &  17.00 &  21.27 $\pm$   0.11 &    0.152 $\pm$    0.007 & 21.13 $\pm$ 0.07 \\
 1022 & 357.38800 &  27.91175 & Sep2010 & 4 &  -1.46 &   9.06 &  22.53 $\pm$   0.24 &    0.055 $\pm$    0.006 &   \\
 1023 &  17.12467 &  24.68193 & Sep2010 & 4 &  -0.55 &  13.64 &  20.81 $\pm$   0.04 &    0.053 $\pm$    0.006 & 20.77 $\pm$ 0.06 \\
 1024 &  17.17512 &  24.68174 & Sep2010 & 4 &  -0.59 &   9.91 &  21.75 $\pm$   0.07 &    0.034 $\pm$    0.004 & 21.91 $\pm$ 0.10 \\
 1025 & 331.97791 &  27.02233 & Nov2010 & 4 &  -0.42 &   6.76 &  20.30 $\pm$   0.05 &    0.032 $\pm$    0.007 &   \\
 1026 & 332.20645 &  24.72709 & Nov2010 & 4 &  -1.17 &  17.20 &  21.23 $\pm$   0.05 &    0.071 $\pm$    0.004 & 21.87 $\pm$ 0.18 \\
 1027 & 332.27611 &  24.61840 & Nov2010 & 4 &  -0.46 &   7.05 &  21.30 $\pm$   0.07 &    0.028 $\pm$    0.005 & 21.94 $\pm$ 0.08 \\
 1028 & 332.53514 &  25.44871 & Nov2010 & 4 &  -1.04 &  10.36 &  21.55 $\pm$   0.12 &    0.062 $\pm$    0.006 & 22.15 $\pm$ 0.11 \\
 1029 & 334.34985 &  27.68723 & Nov2010 & 4 &  -0.45 &   5.71 &  21.97 $\pm$   0.07 &    0.019 $\pm$    0.005 & 22.12 $\pm$ 0.14 \\
 1030 & 334.51783 &  27.55964 & Nov2010 & 4 &  -1.33 &  11.42 &  23.14 $\pm$   0.22 &    0.033 $\pm$    0.004 &   \\
\\
 1031 & 350.37149 &  24.25764 & Nov2010 & 4 &  -0.91 &  21.42 &  21.68 $\pm$   0.05 &    0.045 $\pm$    0.003 & 21.90 $\pm$ 0.09 \\
 1032 & 351.29051 &  25.85756 & Nov2010 & 4 &  -0.69 &   8.87 &  20.90 $\pm$   0.10 &    0.050 $\pm$    0.006 & 21.07 $\pm$ 0.12 \\
 1033 & 351.68935 &  25.68903 & Nov2010 & 4 &  -1.21 &  20.38 &  22.72 $\pm$   0.12 &    0.045 $\pm$    0.004 & 22.25 $\pm$ 0.19 \\
 1034 & 352.94941 &  25.83481 & Nov2010 & 4 &  -0.74 &  47.50 &  19.11 $\pm$   0.02 &    0.274 $\pm$    0.006 & 19.24 $\pm$ 0.03 \\
 1035 & 354.34425 &  25.64395 & Nov2010 & 4 &  -1.98 &   8.47 &  23.51 $\pm$   0.43 &    0.046 $\pm$    0.004 &   \\
 1036 & 354.49840 &  25.69983 & Nov2010 & 4 &  -1.93 &   9.46 &  22.71 $\pm$   0.27 &    0.051 $\pm$    0.004 &   \\
 1037 & 357.03530 &  27.60221 & Nov2010 & 4 &  -0.65 &  15.31 &  20.87 $\pm$   0.06 &    0.057 $\pm$    0.006 & 20.76 $\pm$ 0.07 \\
 1038 & 357.31563 &  25.53299 & Nov2010 & 4 &  -1.51 &  16.02 &  21.68 $\pm$   0.14 &    0.164 $\pm$    0.008 &   \\
 1039 & 358.17888 &  27.30178 & Nov2010 & 4 &  -1.08 &   5.74 &  22.26 $\pm$   0.17 &    0.016 $\pm$    0.004 &   \\
 1040 & 358.24607 &  27.29628 & Nov2010 & 4 &  -0.89 &  10.03 &  22.06 $\pm$   0.14 &    0.042 $\pm$    0.006 & 22.18 $\pm$ 0.14 \\
 \enddata
\tablecomments{Table~\ref{tab:catalog} is published in its entirety in the machine-readable format.  A portion is shown here for guidance regarding its form and content.}
\end{deluxetable*}

After the analysis is completed on the data from each of the AHA observing runs, every \hadot{} candidate in the final list is assigned a unique \hadot{} number.  Following the convention established in K12, the \hadots{} are ordered by increasing right ascension within each observing run.  The observing runs are ordered chronologically, with the \hadot{} numbering sequence proceeding continuously from one run to the next.  To avoid any confusion with the numbering sequence established for the 0.9~m \hadot{} catalogs, the 2.1~m \hadot{} numbers start with 1001.  Hence, the first list of 24 \hadot{} candidates from the September 2010 observing run are numbered 1001 through 1024, the \hadot{} numbers for the 26 ELGs discovered in the November 2010 data run from 1025 to 1050, and so on.  

Our fourth catalog of \hadots{} is presented in Table~\ref{tab:catalog}.   The table includes, for each \hadot{}, its identifying \hadot{} number (column 1), SDSS Right Ascension and Declination (epoch J2000) (columns 2 and 3), the observing run from which the AHA data originates (column 4; see also Table~\ref{tab:obsruns}), and the narrow-band filter used for the observation of that field (column 5, see Figure~\ref{fig:oldhafilts}).  Columns 6 and 7 give the magnitude difference $\Delta m$ and $ratio$ as defined in Section \ref{sec:methodology}.  These represent two of the primary selection parameters used in creating the \hadots{} survey.  The measured R-band magnitude and its associated error is given in column 8, while column 9 gives the measured narrow-band line flux and its error.  These measured quantities are derived directly from our survey images.  Finally, column 10 lists the SDSS r-band magnitude and its error for objects where this has been measured.

We make a few notes about the data presented in Table~\ref{tab:catalog}.  The decision to use SDSS coordinates as opposed to our own astrometry was based on a direct comparison of the two sets of positional data.  The accuracy of our astrometry is, in general, typically quite good.  The median positional offset between our astrometry and SDSS positions is 1.02 arcsec for the full sample.  However, we found that while the positional agreement is good at the centers of our images, it tends to be less reliable near the edges of our frames.  Given the wide-field accuracy of the SDSS astrometry, we have opted to adopt it whenever possible.  In cases where the \hadot{} is located outside the footprint of the SDSS (nine galaxies), the coordinates are those derived from our analysis.  

We include the SDSS r-band magnitudes for those \hadots{} that match with an object in the SDSS photometric catalog (380 objects).   In general, there is good agreement between the SDSS and \hadots{} photometry for objects with R $<$ 22 (with some notable exceptions).  The median SDSS r magnitudes are systematically 0.10-0.15 magnitudes {\it fainter} than our R-band measurements due to well known offsets in the two photometric systems \citep[e.g.,][]{jordi2006}.  For R $>$ 22.5, the photometric errors in both measurements get quite large (typically larger than 0.2 magnitudes).  As expected, the R-band magnitudes for the 65 \hadots{} that were not cataloged by SDSS are substantially fainter than those that are: $\langle$R$\rangle_{In\ SDSS}$ = 21.34, compared with $\langle$R$\rangle_{Not\ in\ SDSS}$ = 22.32.

\begin{figure}[t]
    \centering
    \includegraphics[scale = 0.59]{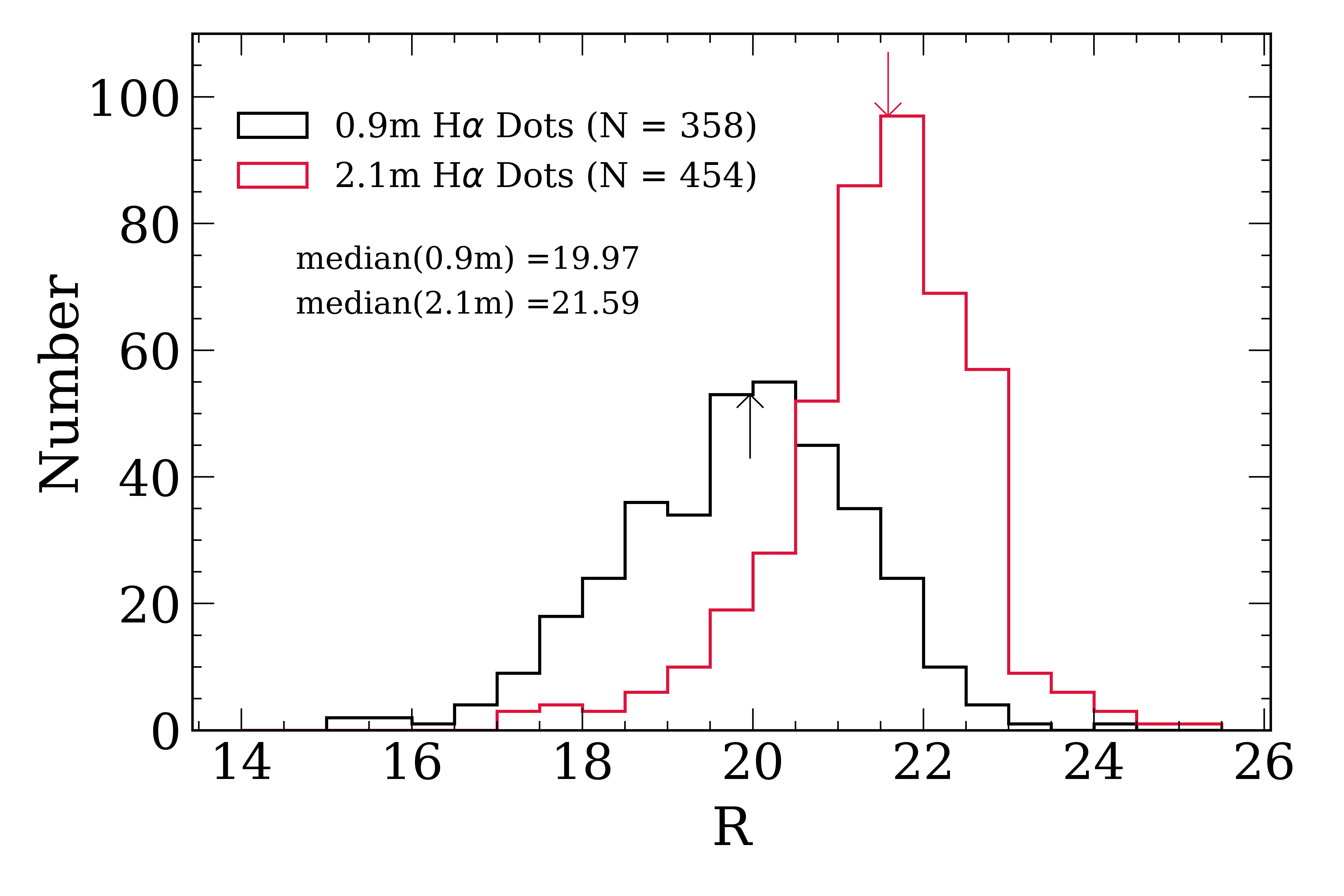}
    \caption{\footnotesize Histogram of the R-band magnitudes for all 454 2.1~m \hadots{} (red histogram).   For comparison we plot the corresponding histogram for the full sample of 0.9~m detected \hadots{} (black histogram).  The median values for the two distributions are given in the legend and are indicated by the two arrows. The median apparent magnitude for the 2.1~m \hadots{} is seen to be 1.62 magnitudes fainter (factor of 4.45$\times$ fainter).}
    \label{fig:JohnsonRmagcomparison}
\end{figure}

Figure~\ref{fig:JohnsonRmagcomparison} presents a histogram of the R-band magnitudes for the full sample of 454 \hadots{} listed in Table~\ref{tab:catalog} (red histogram).  The brightest object in the catalog is \hadot{} 1144, with R = 17.07, while the faintest object, \hadot{} 1117, has R = 25.35.  For comparison, the corresponding histogram for the entire catalog of \hadots{} discovered using the 0.9~m telescope \citep[K12,][plus additional objects from the upcoming third catalog list]{hadots2} is shown in black.  

The median R magnitude in the 2.1~m catalog is 21.59, 1.62 magnitudes fainter than the median R magnitude found in the 0.9~m catalog.  This corresponds to a factor of 4.45 in brightness.  The two sample medians are indicated with arrows in the figure.  The relative photometric depths of the two samples of ELGs is very similar to the ratio of the surface density of objects discussed above, and is consistent with the expectations based on the ratio of the light-collecting areas of the two telescopes of (84 inches/37 inches)$^2$ = 5.15.

{}

\begin{figure}[]
    \centering
    \includegraphics[scale = 0.58]{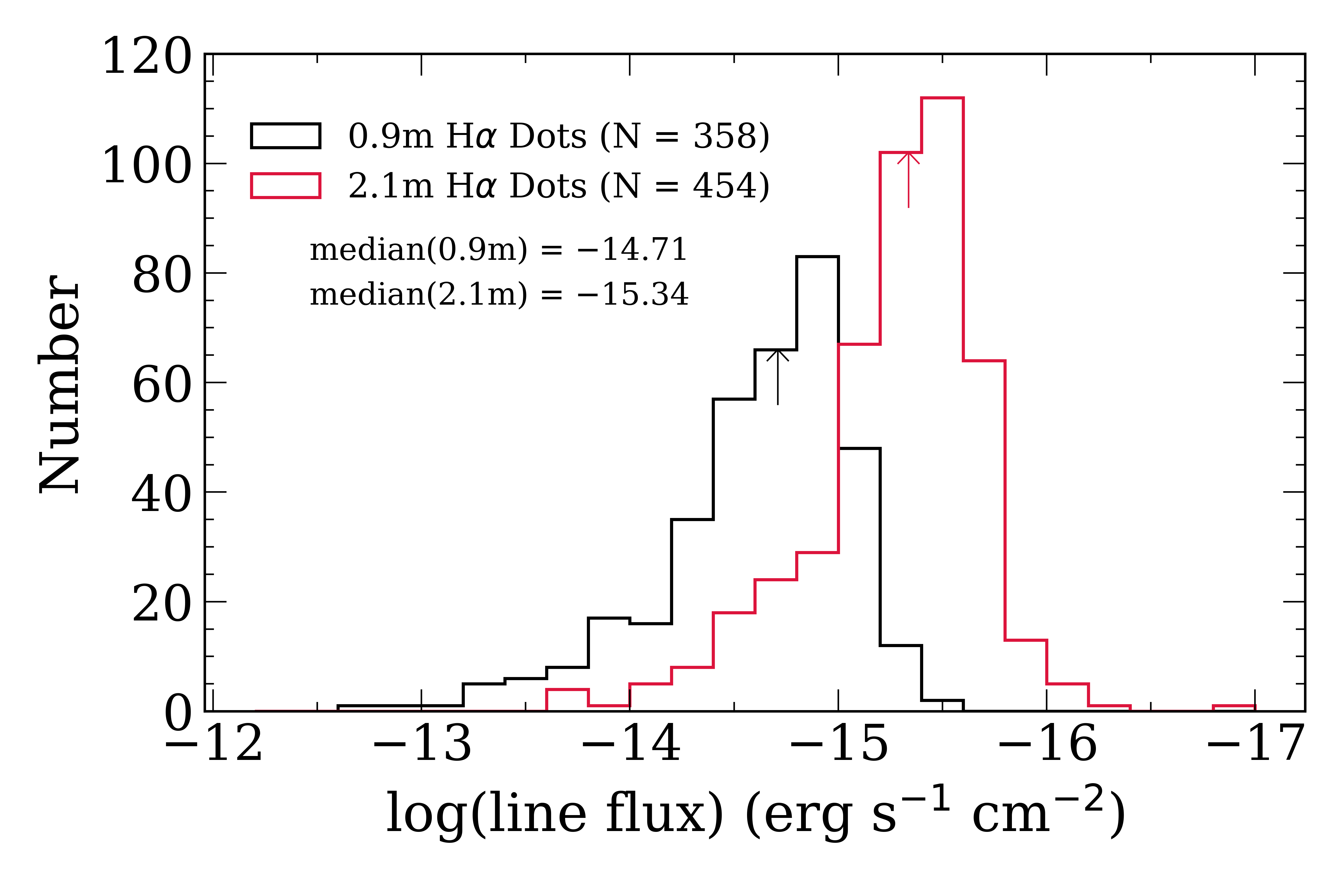}
    \caption{\footnotesize Histogram of the emission-line flux measured in our narrow-band images for the full set of 454 2.1~m \hadots{} (red histogram).   For comparison we plot the corresponding histogram for the entire sample of 358 0.9~m detected \hadots{} (black histogram).  The median values for the two distributions are given in the legend and are indicated by the two arrows. The median line flux for the 2.1~m \hadots{} is seen to be 0.63 dex fainter (factor of 4.27$\times$ fainter).}
    \label{fig:linefluxhistogram}
\end{figure}{}

While the distribution of apparent magnitudes is a useful indicator of the depth of the \hadots{} survey, this is a sample of galaxies detected based on the strength of their emission lines located within the bandpasses of our narrow-band filters.  Hence, the proper way to evaluate the depth of the survey is by examination of the distribution of emission-line fluxes.   This distribution is shown in Figure~\ref{fig:linefluxhistogram}.  

As with Figure~\ref{fig:JohnsonRmagcomparison}, we plot line-flux histograms for both the 2.1~m \hadots{} from the current study (red histogram) as well as the sample of \hadots{} detected in the 0.9~m data (black histogram) in Figure~\ref{fig:linefluxhistogram}.  The median values of the two distributions are indicated in the legend, and denoted by arrows in the plot.  Once again, the 2.1~m \hadots{} sample is seen to be substantially deeper than that found with the 0.9~m telescope.  The median line flux found for the 2.1~m data is 4.57 $\times$ 10$^{-16}$ erg s$^{-1}$ cm$^{-2}$, which is a factor of 4.27 times fainter than the median for the 0.9~m \hadots{}.  The flux distribution of the current sample peaks at log(flux) = $-$15.5 (3.2 $\times$ 10$^{-16}$ erg s$^{-1}$ cm$^{-2}$), beyond which the distribution falls off rapidly.  We adopt this value as the approximate line-flux completeness limit for the 2.1~m \hadots{} sample.

\section{Discussion} \label{sec:discussion}

\subsection{Spectroscopic Follow-up of the New \hadots{}}

Previous \hadot{} survey lists \citep[K12,][]{hadots2}  included information from follow-up spectra.  These spectra provide confirmation of the emission-line detection, as well as identifying the line present in the narrow-band images.  Typically these follow-up spectra yield accurate redshifts and emission-line ratios that allow for the determination of the activity type of each ELG (e.g., star forming vs. AGN).  A similar spectroscopic campaign for objects in the current \hadot{} catalog was not possible, in part because of the larger number of objects, and in part because the objects are, on average, significantly fainter than those in the previous catalogs.  Hence, we are presenting our latest and deepest list of \hadots{} without benefit of spectroscopic confirmations.

During our cross-matching with the SDSS, we noted a small number of \hadots{} with SDSS spectra.  A search of Data Release 16 \citep{sdss16} reveals that 17 \hadots{} from the current survey list (3.7\% of the total) have existing spectra in SDSS.  These objects are listed in Table~\ref{tab:sdssspec}.  There are four \hadots{} with low redshifts where the emission line in our narrow-band filter was H$\alpha$ (based on our inspection of the SDSS spectra).  These four objects are all brighter than R = 18.0, and are among the brightest of the \hadots{} in the current sample (e.g., Figure~\ref{fig:JohnsonRmagcomparison}).  One of these is H$\alpha$ Dot 1144, the brightest object in our catalog.  All four of these were observed as part of the legacy SDSS redshift survey \citep[e.g.,][]{strauss2002}.  The remaining \hadots{} with SDSS spectra were observed as part of the BOSS \citep{boss} or eBOSS \citep{eboss} projects.  One of these, \hadot{} 1021, was detected via strong [\ion{O}{3}]$\lambda$5007 being located in our survey filter.  This object has a spectrum similar to the Green Pea galaxies \citep{gp, brunker2020}.  All of the remaining 12 \hadots{} with SDSS spectra are QSOs.  Of these, nine are detected due to redshifted Mg II emission, one via \ion{C}{3}], and two via \ion{C}{4}.  Table~\ref{tab:sdssspec} lists the \hadot{} number, the emission line detected in our narrow-band filter, the spectroscopic redshift, and the activity type (either QSO or star-forming galaxy (SFG)) for each object.

\begin{deluxetable}{cccc}[t]
\tabletypesize{\small}
\tablewidth{0pt}
\tablecaption{H$\alpha$ Dots with SDSS Spectra\label{tab:sdssspec}}

\tablehead{
 \colhead{H$\alpha$ Dot \#} &   \colhead{Detected Line} &   \colhead{z} &    \colhead{ELG Type}   \\
 (1)  & (2)  & (3)  & (4) 
}
\startdata
 1011 & \ion{Mg}{2} $\lambda$2798 &  1.3955 & QSO \\
 1021 & [\ion{O}{3}] $\lambda$5007  & 0.3406 & SFG \\
 1110 & H$\alpha$ & 0.0165  & SFG \\ 
 1142 & H$\alpha$ & 0.0241  & SFG \\
 1144 & H$\alpha$ & 0.0239  & SFG \\
 1192 & \ion{Mg}{2} $\lambda$2798 & 1.4085 & QSO \\ 
 1193 & \ion{Mg}{2} $\lambda$2798 & 1.3925 & QSO \\
 1214 & \ion{Mg}{2} $\lambda$2798 & 1.3858 & QSO \\ 
 1216 & \ion{C}{4} $\lambda$1549 & 3.2969 & QSO \\ 
 1217 & \ion{C}{4} $\lambda$1549 & 3.3161 & QSO \\
 1220 & \ion{Mg}{2} $\lambda$2798 & 1.3930 & QSO \\ 
 1260 & \ion{C}{3}] $\lambda$1909 & 2.4814 & QSO \\
 1329 & \ion{Mg}{2} $\lambda$2798 & 1.3896 & QSO \\
 1341 & \ion{Mg}{2} $\lambda$2798 & 1.3803 & QSO \\ 
 1356 & \ion{Mg}{2} $\lambda$2798 & 1.3837 & QSO \\
 1358 & \ion{Mg}{2} $\lambda$2798 & 1.3717 & QSO \\
 1392 & H$\alpha$ & 0.0214  & SFG
\enddata

\end{deluxetable}

The distribution of redshifts for the 17 \hadots{} from the current catalog with SDSS spectra is dramatically different from the one for the 0.9 m Dots \citep[e.g.,][]{hadots2}.  The latter sample is dominated by objects detected via their H$\alpha$ or [\ion{O}{3}] lines.  Specifically, 92\% of the \hadots{} from the first two catalogs were detected either by H$\alpha$ (55\%) or [\ion{O}{3}]$\lambda$5007 (37\%).  The remaining galaxies were discovered either by the [\ion{O}{2}]$\lambda$3727 doublet (2\%) or one of the UV lines common to QSO spectra (6\%).

The fact that the ``emission-line detection function" for the \hadots{} found with the KPNO 2.1 m telescope is so different from the one observed for the previous \hadot{} survey lists (WIYN 0.9 m component) should be no surprise.   Many of the objects with SDSS spectra were pre-selected for the BOSS and eBOSS surveys as having broad-band colors consistent with QSOs \citep[e.g.,][]{boss, eboss}.  Hence, the large percentage of QSOs is no accident.  However, even before seeing the SDSS spectra, we were anticipating that the new catalog of \hadots{} would be different.  The increased depth of the 2.1 m sample compared to the 0.9 m Dots, coupled with the fixed redshift range accessible for each emission line, implies that the fainter \hadots{} should preferentially be objects selected at higher redshifts rather than being lower luminosity versions of the the objects found with the 0.9 m telescope.  We expect a higher portion of [\ion{O}{3}]-detected Dots in the current catalog than were present on the first two survey lists, as well as substantially higher numbers of [\ion{O}{2}]-detected galaxies and high redshift QSOs.  As time and telescope resources allow, we will hopefully get the opportunity to test these hypotheses as we obtain follow-up spectra for these new \hadots{}.

\subsection{Applications of the \hadots{}}

As highlighted in the previous survey lists based on sources detected in WIYN 0.9 m images \citep[K12,][]{hadots2}, the \hadots{} have a number of interesting science applications.  These include studying large samples of dwarf star-forming galaxies, including Blue Compact Dwarfs (BCDs), and the detection of strong [\ion{O}{3}] emitters like Green Peas and Seyfert 2 galaxies.  The lack of existing follow-up spectroscopy for the current sample of \hadots{} naturally limits its immediate impact on addressing relevant science questions.  Nonetheless, we draw attention to the high-impact scientific applications these objects can be used to address.

\subsubsection{Low Luminosity Star-forming Galaxies} 
As outlined above, we anticipate that a lower percentage of the \hadots{} in the current catalog will be low redshift H$\alpha$ detections.  Still, we expect that a significant fraction will have been detected via the H$\alpha$ line.  Given the wavelength coverage of the narrow-band filters employed, the resulting redshift range of the H$\alpha$-selected galaxies will be 0.005 -- 0.026 (velocity range 1460 -- 7810 km/s).  This redshift range, coupled with the apparent magnitude distribution of the current sample (e.g., Figure~\ref{fig:JohnsonRmagcomparison}), implies that the H$\alpha$-detected \hadots{} will all be dwarf star-forming systems \citep[e.g., see Figure 8 in][]{hadots2}.  Since this catalog of galaxies represents a statistically complete, line-flux-limited sample, it could be used for accurately establishing the volume density of low-luminosity star-forming galaxies in the local universe.

Since the \hadots{} are pre-selected as possessing strong emission lines, this sample of dwarf galaxies will also be ideal for measuring metal abundances for a large, statistically-complete sample of objects.  A similar study is currently underway utilizing the brighter \hadots{} from K12 and \citet{hadots2} (A. Hirschauer, in preparation).  An important application of the current catalog of objects is that they extend the detection of strong-lined ELGs to substantially fainter magnitudes.   Hence, we expect that the 2.1 m survey of \hadots{} will include lower-luminosity dwarfs than those found in the previous survey lists.   This in turn should result in a larger yield of extremely low abundance systems, similar to HADot 303 = AGC 198691 \citep[a.k.a. the Leoncino Dwarf;][]{alec2016, mcquinn2020}.

\subsubsection{[\ion{O}{3}]-detected Star-forming Galaxies} 
The \hadots{} discovered with the WIYN 0.9 m telescope included a large number of [\ion{O}{3}]-detected galaxies (37\% of the objects in the first two survey lists).  The expectation is that the 2.1 m Dots will include a comparable or somewhat larger fraction of systems detected by the [\ion{O}{3}]$\lambda$5007 line.  In fact, the current catalog might well be dominated by such objects.  The relevant redshift range for detection by [\ion{O}{3}]$\lambda$5007 is z=0.317-0.345.  

The [\ion{O}{3}]-detected systems found in the previous \hadots{} catalogs included a mix of Green Pea-like galaxies \citep{gp, brunker2020} and Seyfert 2 galaxies.  We expect that the current catalog will detect many additional Green Pea candidates.   Additionally, given the increased depth of the current list of \hadots{}, we fully expect that many less extreme [\ion{O}{3}]-selected
star-forming galaxies will also come within reach of detection.  It is well known that the strength of the [\ion{O}{3}]$\lambda$5007 line peaks at metal abundances of $\sim$10\% solar (log(O/H)+12 $\sim$ 7.7).  In the local universe (z $<$ 0.1), actively star-forming galaxies with B-band absolute magnitudes in the range $-$16 $\le$ M$_B$ $\le$ $-$18 are often found in [\ion{O}{3}] line-selected samples such as the UM survey \citep[e.g.,][]{UM1977, UM1981, salzer1989}, the Case survey \citep[e.g.,][]{case82, case83, salzer1995}, and the KISSB survey \citep{salzer2002}.  The depth of the current sample should allow for the detection of this population of objects at the higher redshifts probed by the [\ion{O}{3}]$\lambda$5007 line with our filters.

A key attribute of both the Green Pea galaxies and the intermediate luminosity [\ion{O}{3}]-detected SFGs is that they very often exhibit spectra with such strong emission lines that the temperature-sensitive [\ion{O}{3}]$\lambda$4363 auroral line is present in their follow-up spectra.  Hence, we expect that the [\ion{O}{3}]-detected \hadots{} in the current list will include dozens of sources from which accurate direct abundances will be measurable.  

\subsubsection{ [\ion{O}{3}]-detected Seyfert 2 Galaxies} 
The additional depth of the current \hadot{} catalog will likely result in a deeper and presumably more comprehensive sample of [\ion{O}{3}]-selected Seyfert 2 galaxies in the redshift window z=0.317-0.345.  While the previous \hadots{} list include a significant number of Seyfert 2s (11\% of the sample overall, and 30\% of the [\ion{O}{3}]-detected Dots), they tend to be objects with extreme spectral characteristics.  The Seyfert 2s included in the previous \hadots{} catalogs tend to exhibit very high [\ion{O}{3}]/H$\beta$ ratios; they are nearly all very high-excitation objects (see Figure 7 in \citet{hadots2}).  Because the great depth of the current survey, the Seyfert 2s cataloged in the current paper should include both the high-excitation objects as well as many with lower [\ion{O}{3}]/H$\beta$ ratios.  Overall we expect an even higher percentage of Seyfert 2s compared to the previous lists.  We also expect that this new sample of Seyferts will be more representative, rather than being biased toward the more extreme examples.

Once again, the line-flux limited nature of the \hadots{} survey method will allow us to generate an accurate volume density for Seyfert 2 population in the redshift window covered by our filters.  The strong-lined nature of the Seyfert 2 sample will also allow for the study of the metallicity of the AGN at these redshifts (which represents a lookback time of $\sim$3.7 Gyr).  A preliminary analysis of the [\ion{O}{3}]-detected Seyfert 2s from the previous survey catalogs has indicated the possibility of a modest drop in the average metal abundance compared to low redshift counterparts (D. Carr, in preparation).

\subsubsection{ [\ion{O}{2}]-detected SFGs and AGN} 
Another expectation of the current list of \hadots{} is that it will include a higher proportion of [\ion{O}{2}]-detected SFGs and AGN at z = 0.770-0.807.  This population of ELGs is just barely detectable with the 0.9 m telescope portion of the \hadots{} survey.  Only three [\ion{O}{2}]-detected galaxies are included in the first two survey lists, and these are all found in the fainter portion of the sample (average line flux of 1.3 $\times$ 10$^{-15}$ ergs/s/cm$^2$).  The increased depth of the 2.1 m survey list should result in a substantial increase in the number of [\ion{O}{2}]-selected ELGs.  This will allow the survey to probe the star-forming and AGN galaxy populations at these cosmologically interesting distances (lookback times of $\sim$6.8 Gyr, or 50\% the age of the universe).

\subsubsection{ QSOs} 
Finally, we mention the rather obvious presence of numerous QSOs within the current \hadots{} catalog.  While our survey is not capable of producing a comprehensive QSO survey, it does detect substantial numbers of quasars in specific, narrow redshift windows: z = 1.357-1.408 for the \ion{Mg}{2} $\lambda$2798 line, z = 2.454-2.528 for \ion{C}{3}] $\lambda$1909, z = 3.257-3.347 for \ion{C}{4} $\lambda$1549, and z = 4.428-4.543 for Ly$\alpha$ $\lambda$1215.  If detected in sufficient numbers, the \hadots{} QSOs could provide accurate ``hard points" for their volume densities in these redshift windows.  While it is clear that the large fraction of QSOs among the \hadots{} with available SDSS spectra is a selection effect, these spectra provide a glimpse of the potential science applications of the line-flux limited \hadots{} survey for probing the QSO population at a range of redshifts.

\section{Summary \& Conclusions} \label{sec:summary}

We present the latest list of \hadots{}, based on images obtained for the ALFALFA H$\alpha$ project \citep{vansistine16}.  \hadots{} are compact emission-line sources detected serendipitously in narrow-band images.  Previous survey catalogs have presented lists of \hadots{} detected in images obtained with the WIYN 0.9 m telescope \citep[K12,][]{hadots2}.  Our new list of \hadots{} has been created by analyzing 611 ALFALFA H$\alpha$ fields observed with the KPNO 2.1 m telescope.

The current \hadot{} catalog contains 454 unique \hadots{}.  All the new \hadots{} were identified in  2.1 m images using the same software packages developed for the previous \hadots{} catalogs.  Hence, the only significant difference with the previous survey lists is in the depth of the sample.  The 2.1 m \hadots{} survey is sensitive to fainter objects, detecting sources with a median apparent R magnitude of 21.59 and median line fluxes of 4.57 $\times$ 10$^{-16}$ erg s$^{-1}$ cm$^{-2}$.  In both metrics, the current survey list probes a factor of $\sim$4.4 times fainter than the 0.9 m \hadots{} catalogs.  The approximate emission-line flux completeness limit of the current sample is 3 $\times$ 10$^{-16}$ erg s$^{-1}$ cm$^{-2}$.

While the previous \hadots{} catalogs included information from follow-up spectroscopy, we do not have corresponding spectral data for the current list of ELGs.  We speculate that the additional depth of the \hadots{} list generated using 2.1 m telescope images will result in a significantly different mix of objects being discovered, relative to the previous catalogs.  While we expect that the current catalog includes numerous low-luminosity star-forming dwarf galaxies detected via their H$\alpha$ lines, we expect that this population will account for a much smaller fraction of the overall ELG catalog when compared to the lists generated from the 0.9 m data (where the H$\alpha$-detected fraction was 55\%).  We anticipate large fractions of the current catalog will be found to have been detected via their [\ion{O}{3}] or [\ion{O}{2}] lines.  Plans for carrying out follow-up spectroscopy of the 2.1 m \hadots{} are being formulated.

\acknowledgments

We would like to thank the anonymous referee who made a number of suggestions that have improved the paper.  We gratefully acknowledge the financial support of the College of Arts and Sciences and the Department of Astronomy at Indiana University, which helped make this ongoing undergraduate research project possible.  The H$\alpha$ Dots survey project is based on data obtained for the ALFALFA H$\alpha$ project, which was carried out with the support of a National Science Foundation grant to JJS (NSF-AST-0823801).  We would also like to acknowledge the Maria Mitchell Association, which provided a summer research internship to RR as part of their NSF-funded REU program (with JJS serving as an associate mentor).

This project made use of Sloan Digital Sky Survey data.  Funding for the SDSS and SDSS-II has been provided by the Alfred P. Sloan Foundation, the Participating Institutions, the National Science Foundation, the U.S. Department of Energy, the National Aeronautics and Space Administration, the Japanese Monbukagakusho, the Max Planck Society, and the Higher Education Funding Council for England. The SDSS Web Site is http://www.sdss.org/.  The SDSS is managed by the Astrophysical Research Consortium for the Participating Institutions. The Participating Institutions are the American Museum of Natural History, Astrophysical Institute Potsdam, University of Basel, University of Cambridge, Case Western Reserve University, University of Chicago, Drexel University, Fermilab, the Institute for Advanced Study, the Japan Participation Group, Johns Hopkins University, the Joint Institute for Nuclear Astrophysics, the Kavli Institute for Particle Astrophysics and Cosmology, the Korean Scientist Group, the Chinese Academy of Sciences (LAMOST), Los Alamos National Laboratory, the Max-Planck-Institute for Astronomy (MPIA), the Max-Planck-Institute for Astrophysics (MPA), New Mexico State University, Ohio State University, University of Pittsburgh, University of Portsmouth, Princeton University, the United States Naval Observatory, and the University of Washington.

\vspace{5mm}
\facilities{KPNO:2.1m}

\software{IRAF}


\

\end{document}